\documentstyle[epsfig,12pt]{article}
\newcommand {\ds}   {\displaystyle} 
\newcommand {\be}   {\begin{equation}}
\newcommand {\ba}   {\begin{array}}
\newcommand {\bea}  {\begin{eqnarray}}
\newcommand {\bfi}  {\begin{figure}}
\newcommand {\ee}   {\end{equation}}
\newcommand {\ea}   {\end{array}}
\newcommand {\eea}  {\end{eqnarray}}
\newcommand {\efi}  {\end{figure}}                  
\newcommand {\dV}  {\delta V}
\newcommand {\UNIV}   {Universit\`a }
\begin{document}
%
\title{Scaling laws and intermittency in homogeneous shear flow}
\author{
P\@. Gualtieri
\thanks{Dip. Mecc. Aeron., \UNIV di Roma ``La Sapienza'',
        via Eudossiana 18, 00184, Roma, Italy.},
C\@.M\@. Casciola$^*$
R\@. Benzi
\thanks{AIPA, via Solferino 15, 00185, Roma, on leave of absence from
        Dip. di Fisica, \UNIV di Roma ``Tor Vergata'', Italy.}
G\@. Amati
\thanks{CASPUR, p.le A. Moro 5, 00185 Roma, Italy.},
\&
R\@. Piva$^*$.}
\maketitle
\section{Abstract}
In this paper we discuss the dynamical features of intermittent fluctuations
in homogeneous shear flow turbulence. In this flow the energy cascade is 
strongly modified by the production of turbulent kinetic energy related to 
the presence of vortical structures induced by the shear. By using direct 
numerical simulations, we show that the refined Kolmogorov
similarity is broken and a new form of similarity is observed, in agreement 
to previous results obtained in turbulent boundary layers. As a consequence,
the intermittency of velocity fluctuations increases with respect to 
homogeneous and isotropic turbulence. We find here that the statistical
properties of the energy dissipation are practically unchanged with respect
to homogeneous isotropic conditions, while the increased 
intermittency is entirely captured in terms of the new similarity law.
\newpage
\section{Introduction}

The issue of anomalous scaling laws in turbulence
has been largely addressed for the idealized case of 
homogeneous and isotropic turbulence once it became clear that a purely dimensional
power law \cite{kolm_41} can not be consistent with the intermittent nature of energy 
dissipation \cite{Landau}. The Kolmogorov-Obukhov refined similarity (RKSH) 
\cite{kolm_62} provides the key theoretical point by relating the anomalous correction 
of the scaling exponents of velocity structure functions directly to the statistical 
properties of the dissipation field. The RKSH has been subject to close 
scrutiny by a number of independent investigations making use of both experiments and 
direct numerical simulations and it may be considered a well assessed physical law. 
One of the early difficulty for an accurate evaluation of scaling exponents, 
associated to the existence of too small scaling ranges in moderate Reynolds number 
turbulence, has been recently overcome with the use of the Extended Self-Similarity (ESS) 
\cite{benzi_1}. 
This approach has allowed accurate estimates of the exponents 
from moderate Reynolds number flows  \cite{frisch} and the evaluation of the RKSH, in 
its extended form, from DNS data \cite{benzi_2}.

A more recent research interest in turbulence has been focused on the assessment of RKSH
in non-isotropic and non-homogeneous conditions, namely in flows characterized by strong 
mean shear and anisotropy. The main focus is to understand to what extent 
RKSH may be considered still able to capture, in such conditions, the essential features 
of turbulence dynamics.

In this context recent results for a turbulent channel flow \cite{PF}, have pointed out 
how the classical RKSH holds in the higher part of the logarithmic region. 
This result should not be surprising since the RKSH is established as a consequence of 
the  balance between energy transfer due to non linear interactions and dissipation. 
In the log-layer, where the mean shear is weak, the dynamics of the flow is 
characterized by the inertial energy transfer. In fact, when approaching the wall, the 
shear becomes larger and larger and a clear failure of the RKSH is observed \cite{PF}. 
This result gives a partial answer to the point raised above since it gives evidence 
that, in the near wall region, the homogeneous isotropic behavior is far from being 
recovered. 

At this point a question naturally arises about the existence of an alternative scaling 
law in such conditions. On the basis of physical considerations on the respective role 
of energy production and energy transfer a different Kolmogorov-like scaling law has 
been proposed in \cite{PF} and preliminarily checked against DNS data from the buffer 
region of a turbulent channel flow. The validity of this similarity law has been 
confirmed by a successive experimental work in a turbulent boundary layer.
However, certain aspects of the problem are still not completely clear.
Actually, the new similarity law has been conceived as a direct consequence of the 
presence of a strong shear, but it has been verified only in wall turbulence where, in 
principle, some other features such as the continuous variation of the local shear, the 
non uniform momentum flux across adjacent layers and the suppression of wall normal 
velocity fluctuations may play a quite significant role.

To isolate the effect of the shear we reconsider here the problem in the context of a 
DNS of a homogeneous shear flow in a confined box. In fact, in the rationale of dimensional 
analysis, this flow presents the minimum level of complexity though maintaining the 
essential features we want to address: thus, it is ideal to study the effect of 
a pure shear avoiding other concurrent effects. On the other hand, it allows to exploit 
homogeneity in all spatial directions so that a more complete and accurate statistical 
analysis can be performed, at a level of detail that could never be achieved in the very 
limited buffer region of a wall bounded flow.

As a counterpart, the homogeneous shear flow presents certain drawbacks related to the 
artificial nature of its confinement. To understand to what extent the present flow may 
serve as a prototype for shear-dominated flows, we do analyze in detail 
the dynamics of the coherent vortical structures observed in the numerical simulation. 
This is instrumental to qualify the flow, i.e. its dependence on the aspect ratio of the 
computational box, in view of the use we do of the flow itself for the evaluation of scaling 
laws and their assessment.

Homogeneous shear flows have been investigated in the literature in many 
theoretical, numerical and experimental works. Rogers and Moin \cite{rogers} and Lee 
{\em{et. al.}} \cite{lee} studied the topology and mutual interactions of vorticity
showing that many dynamical features are quite similar to 
those observed in the wall region of a turbulent boundary layer. A more detailed study 
on the same subject has been performed by Kida and Tanaka \cite{kida} who discussed   
the regeneration cycle of the streamwise vortices in a homogeneous shear flow.   
   
Experimentally the papers by Rose \cite{rose}, Champagne Harris and Corrsin 
\cite{champagne}, Tavoularis and Corrsin \cite{tavoularis_1} \cite{tavoularis_2},    
have investigated homogenous shear flows obtained in a wind tunnel, focusing on the local
isotropy of small scale fluctuations. Following the Kolmogorov theory,   
an important question to investigate, which becomes extremely well posed in the case of 
the homogeneous shear flow, concerns the recovery of small scales  local isotropy
in turbulent flows characterized by large scale non isotropic forcing. 
The same kind of question has been addressed by Saddoughi and Veeravalli 
\cite{saddoughi} who investigated local isotropy for the turbulent flow in a logarithmic 
boundary layer.    
   
The issue of local isotropy has been also addressed more recently by Pumir and Shraiman 
\cite{pumir_2} and by Pumir \cite{pumir_3} who performed an extensive analysis of    
numerical simulations of statistically stationary homogeneous shear flows. Interestingly 
enough, most of the statistical properties found by Pumir \cite{pumir_3} are quite close 
to the experimental findings of Garg and Warhaft \cite{garg} and Shen and Warhaft 
\cite{shen}. In the latter experiment, for the first time, an active grid was used in 
order to produce a homogeneous shear flow in a wind tunnel whose integral scale in the 
streamwise direction is rather constant.   
   
Taking full advantage of most of these results, and in order to be able to reconsider
the issue of local isotropy from a different perspective, as we plan to do in the near 
future, our interest here is mainly focused on the relationship between intermittency and the
anisotropy induced by the mean velocity gradient at large scale.  In the context of
boundary layer we have shown how the increase of intermittency may be explained in 
terms of the phenomenology described by the new form of scaling law. In particular,  
as suggested in \cite{PF}, a new  length scale $L_s  = \sqrt{\bar{\epsilon} / S^3}$ enters 
in the  description of the statistical properties of turbulence, where $\bar{\epsilon}$ is 
the mean rate of energy dissipation and $S$ the mean large scale shear. 
For scales smaller than $L_s$ the statistical 
properties of the turbulent fluctuations are similar to those observed in homogenous and 
isotropic flows while for scales larger than $L_s$ the refined  Kolmogorov similarity 
(RKSH) is broken  and intermittency increases. The phenomenological theory developed for 
the turbulent  boundary layers relies upon the mean shear strength $S$ as the only 
parameter which fixes the scale $L_s$ where a new form of RKSH should be observed.    
For this reason, the homogenous shear flow becomes a natural test case to understand whether 
or not the phenomenological theory devised for the  boundary layer is sufficiently 
general to be applied to any turbulent flow.
As we shall discuss in this paper, our numerical simulations of homogenous shear flows 
support rather well the new phenomenology, leading to possible important implications in 
our understanding of shear turbulence.   
   
The paper is organized in the following way. In section 3 we briefly describe how the 
numerical simulations are performed. 
In section 4 we discuss the regeneration cycle of the vortical 
structures observed in the flow while in section 5 we address the intermittency cycle 
which characterizes the dynamical behavior of the system. In section  6, after a short 
review of the phenomenological theory proposed for the turbulent boundary layer, we describe 
the intermittency properties of turbulent fluctuations. In section 7, we investigate the 
anomalous scaling of the energy  dissipation field, and, finally, in section 8 we draw our 
main conclusions.     
\section{Homogeneous shear flow}
We consider here a turbulent flow in a confined box with an imposed mean shear $S$ 
free from boundaries.
The full Navier-Stokes equations are solved (DNS), after decomposing the velocity field 
into mean value and fluctuation
\be
\label{velo_dec}
 \vec v(\vec x,t)= U(y) \vec e_1 + \vec u(\vec x,t) \, ,
\ee
where 
$\vec x \in V =[-\lambda_x,\lambda_x] \times [-\lambda_y,\lambda_y] 
\times [-\lambda_z,\lambda_z]$, $V$ identifies the computational box, $U(y)=S(y+\lambda_y)$ 
is the mean flow and $\vec e_1$ is the unit vector in the streamwise direction $x$. 
The mean gradient $S$ is in the normal direction $y$ while $z$ denotes the spanwise coordinate. 
The Navier Stokes equations are written in terms of velocity fluctuations 
\be
\label{shear_eq}
\left\{
\ba{l}
\ds  \vec\nabla\cdot\vec u=0 \\ \\
\ds \frac{\partial\vec u}{\partial t}=(\vec u {\bf{\times}} \vec \zeta)
             - \vec \nabla (p+\frac{u^{2}}{2})
             +\nu \nabla^{2}\vec u -Sv \vec e_{1}
             -U(y) \frac{\partial \vec u}{\partial x} \, ,
\ea
\right.
\ee
where $\vec \zeta$ is the vorticity, $v$ the normal velocity, $p$ the pressure and $\nu$ the 
kinematic viscosity.
To achieve efficiency and accuracy a Fourier spectral method is advisable to solve the 
initial value problem for eq. (\ref{shear_eq}).
However, due to the mean flow, the intrinsically  non-periodic term 
$U(y) \partial \vec u / \partial x$ appears in the equations. 
Luckily, by using Rogallo's technique \cite{rogallo}, this difficulty is removed by the 
transformation
\be
\label{var_transf}
\left\{
\ba{l}
    \ds  \xi=x-U(y)t \\
    \ds  \eta=y     \\
    \ds  \zeta=z     \\
    \ds  \tau=t.
\ea
\right.
\ee
In the new variables periodic boundary conditions can be enforced in all spatial directions 
allowing efficient pseudo-spectral methods to be used for spatial discretization. 
The time integration is performed using a third order low storage Runge-Kutta method 
\cite{solver} and the nonlinear terms are fully de-aliased by zero padding. 
Since the image of the computational box in physical space gets distorted,  
eq. (\ref{var_transf}), a re-meshing procedure is periodically applied to allow long time 
integrations. 
Using periodicity in the $\xi$ direction, the computational domain is transformed back
into a non skewed domain whenever the plane $y=\lambda_y$ has moved by $2\lambda_x$
in the streamwise direction,  i.e. every $\Delta t_r =\lambda_x / S \lambda_y$.
This procedure may introduce aliasing errors since the dynamics of wave vectors in 
physical space is time dependent \cite{townsend}, as seen from the relation between 
the Fourier transforms in the computational and in the physical space, 
\be
\label{coef_rel}
\hat{u}_i(k_x,k_y,k_z,t)=\hat{\bar{u}}_i(k_{\xi},k_{\eta}-S \tau k_{\xi},k_{\zeta},\tau).
\ee
To avoid aliasing errors, the spectral components with wave numbers outside the 
interval $[-k_{max},k_{max}]$ are set to zero at re-meshing. 
This filtering introduces a characteristic wave number defined as
\be
\label{k_remesh}
  k_{r}=\frac{k_{max}}{\sqrt{1+A^2}}
\ee
where $k_{max}= \pi N_y / 2 \lambda_y$ and $A=\lambda_x / \lambda_y$ is the aspect 
ratio of the computational domain. In order to asses the influence of the filtering,
the aspect ratio of the box and the resolution of the grid were varied. 
In all cases $k_r$ was larger than the dissipation wavenumber and no appreciable alteration
of the dynamics was observed.

A crucial point for our successive analysis is the ability of the system to reach statistical 
stationarity. 
Under this respect, our computations entirely confirm the results of Pumir and Shraiman 
\cite{pumir_2} and Pumir \cite{pumir_3}.
Actually the evolution of the turbulent kinetic energy is described by the equation
\be
\label{energy_eq}
\frac{\partial}{\partial t} [\frac{u^2}{2}] + S[u v] = -\nu [ \zeta^2]
\ee
where the square brackets denote spatial average. 
This equation suggests a possible balance between the production $S[u v]$ and the 
dissipation $\nu [\zeta^2]$.
This balance is not achieved for the instantaneous fields, as shown in figure 1
which reports the history of turbulent kinetic energy in our longest 
calculation. 
Large fluctuations in energy ($42 \%$) and enstrophy ($50 \%$) are apparent in the 
pseudo-cyclic behavior shown in the figure. The fluctuation level in this system
is quite huge, expecially when compared to that of forced homogeneous isotropic
turbulence, which is limited to only a few percent of the rms value ($5-6 \%$) for both kinetic
energy and enstrophy.
In fact, the pseudo-cyclic behavior corresponds to statistically stationary conditions, as 
follows from time averages performed over time periods much larger than the typical length 
of the cycle.

Clearly, DNS enables to achieve a good level of homogeneity, as shown in figure 2.
In particular both the turbulent kinetic energy and the Reynolds stresses are substantially
constant in the $y$ direction.  

In order to check the dynamics of the velocity gradients, the probability density
function of $\partial u / \partial x$ and of $\partial u / \partial y$ have been computed,
figure 3 and 4. They are often used to asses the small 
scale dynamics of the flow and to characterize its degree of anisotropy.
The computed values of skewness and flatness are in good agreement with the experimental 
results at $Re_{\lambda} \sim 100$ of Shen and Warhaft \cite{shen} and of Ferchichi and
Tavoularis \cite{ferchichi} at $Re_{\lambda}=140$.
\section{Regeneration cycle of vortical structures}
The energy fluctuations discussed in the previous section correspond to a regeneration 
cycle of the vortical structures.  Most of the previous numerical simulations have described
the main features of coherent structures and the vorticity statistics for the early
stages of development of the flow.
In particular, Rogers $\&$ Moin \cite{rogers} showed typical hairpin vortex structures 
remarkably similar to those observed in wall bounded flows.
Lee {\em{et. all.}} \cite{lee} discussed streamwise vortices and high and low speed streaks 
commenting on the similarities with the buffer region of wall bounded flows. 
In a more recent work Kida $\&$ Tanaka \cite{kida} proposed a regeneration mechanism 
for the  streamwise vortices and pointed out the role of vortex sheet instability in the
formation of new vortices.
The present analysis deals with the same issues, but it is focused on the statistical steady 
state regime of the flow. 

The phenomenology is better understood by considering a few snapshots along the history
of the turbulent kinetic energy and of the Reynolds stresses. 
In figure 5 the large energy bursts already noticed in the previous 
section are clearly correlated to large negative values of the Reynolds stresses.
The bursts are induced by large rates of energy injection from the mean flow associated with
the presence of streamwise vortices.
Figure 6, corresponding to an instant at the beginning of a stage of
energy growth, actually shows a large population of quasi-streamwise vortices, here
visualized through the discriminant of the velocity gradient \cite{chong}.
During the successive phase of energy growth the streamwise vortices, by interacting with 
the mean velocity field, give rise to instantaneous profiles characterized by the 
typical ramp and cliff pattern. The pdf of $\partial u / \partial y$ is a suitable tool
to characterize statistically the ramp and cliff structures, figure 4. 
The skewness of $0.82$ gives reason of intense positive events of $\partial u / \partial y$ 
much more probable than negative ones.
As for the passive scalar, the streamwise velocity recovers the mean gradient through ramps 
followed by cliffs \cite{pumir_1}. 
The ramps correspond, on average, to streamwise velocity increasing with $y$, hence
a resulting negative spanwise vorticity is expected, on average.
The pdf of the spanwise vorticity, figure 7, confirms this analysis.

Ramp and cliffs are associated with thin regions of (negative) spanwise vorticity, i.e.
vortex sheets, as shown in figure 8.
The sheets become unstable and eventually roll up into spanwise vortices through the classical  
Kelvin-Helmholtz mechanism, see figure 9 where spanwise vortices are 
systematically located in the regions where the roll-up process is occurring.
Vortex sheets and spanwise vortices are clearly responsible for the observed energy bursts
by producing large negative Reynolds stress events.

Immediately after each burst, the non-linear interactions are enhanced and the original 
vortex structures lose their order resulting into a randomized  vorticity field.
This phase, shown in figure 10, reduces the level of anisotropy of the flow. 
However, soon afterwards, the shear term of the vorticity equation enforces a mean orientation 
to the structures, see figure 11, corresponding to a minimum for the energy, 
just before the occurrence of the successive burst. 
The vortical structures are now getting more and more aligned with the streamwise direction 
and a new cycle starts.

The regeneration cycle just analyzed here for the statistical steady state phase 
is very similar to that described by Kida $\&$ Tanaka \cite{kida} for the early stages of 
evolution of the flow from isotropic initial conditions, suggesting that the operating
mechanisms are substantially identical.
\section{Intermittency cycle}
The increase of the turbulent kinetic energy due to the streamwise vortices can be 
understood, basically, in terms of the linear lift-up mechanism and the related 
transient growth.
A non linear mechanism is required instead to explain the saturation and the break-down 
of the ordered system of vortices. 
In spectral space, as already discussed by Pumir \cite{pumir_3}, the Fourier mode 
$(0,0,\pm 1)$ gives the leading contribution to the energy growth. 
From the linearized equation for this  mode,  
\be
\label{linear_ns}
\left\{
\ba{l}
\ds  \frac{d \hat{u}}{dt}=-S \hat{v} - \nu \hat{u}\\ \\
\ds \frac{d \hat{v}}{dt}= - \nu \hat{v} ,
\ea
\right. 
\ee
a growing amplitude is expected whenever 
$S[Re(\hat{u}) Re(\hat{v}) + Im(\hat{u}) Im(\hat{v}) ] < 0$. 
As soon as the energy grows appreciably, the energy drained by the interactions with the 
other modes in favor of the smaller scales quite rapidly saturates its amplitude.
Hence, the characteristic frequency of the intermittency cycle is determined by the 
dynamical balance between growth of the basic mode and energy transfer. 

In principle, if the mode $(0,0,\pm 1)$ were isolated from the others during its growth, the
energy due to the interaction with the mean shear would be almost entirely 
found in this mode. Then, the characteristic time of the system would 
essentially coincide with the eddy turn over time of the basic mode, hence to the spanwise 
size of the box $L_z$. 

In fact, the leading mode evolves in presence of many others. 
In this case the saturation time, i.e. the time of the effective activation of the nonlinear 
energy transfer, is related to the separation, in wavenumber space, between the leading 
mode and the small scales modes. 
The activation time becomes shorter and shorter as the distribution of energy among all 
Fourier modes approaches a continuous distribution. For the statistical steady state regime, 
in particular, the non linear energy transfer is crucial from the very beginning of the growth.
Actually the energy spectrum in correspondence of the large bursting phenomena, 
figure 12, shows initialy a growth which, though dominated by the
mode $(0,0,\pm 1)$, is spread all over the modes in the first decade. The evolution of the
spectra manifest a strong energy transfer which though occurring during the entire cycle, is 
clearly prevailing in the phase of energy decrease.
Hence the bursting frequency can not be estimated from the isolated dynamics of the 
sole $(0,0,\pm 1)$ mode, and, as a consequence, there is no {\sl a priori} reason for a strong 
dependence of the bursting period on the size of the box.

In view of a quantitative analysis of this issue, let $E_U$ and $E_u$ denote the kinetic 
energy of the basic flow and of the fluctuations,  respectively. 
Clearly, $E_u$ is a function of time.
Two quite different cases can be distinguished, namely
\begin{itemize}
\item[$a)$] $E_u \geq E_U$ at $t=0$
\item[$b)$] $E_u \ll  E_U$ at $t=0$ \, .
\end{itemize}                                                
Case b) corresponds to a problem of transition to turbulence from small, though finite, 
disturbances and will not be considered here in further detail. We are presently 
interested in case a), which is more appropriate to describe statistical stationarity.  
Here $t=0$ should be understood as an arbitrary instant of time along the cyclic 
evolution of the system.

Given the correlation function $C(\tau) = <E_u(t + \tau) E_u(t)>$, let us consider the 
correlation time, $\tau_C$, defined, e.g., as the smallest $\tau$ such that 
$C(\tau_C)=1/2 C(0)$.
Following the discussion of section 4, $\tau_C$ corresponds, roughly, to the characteristic 
time of production of large scale velocity$/$vorticity instability which is the primary 
forcing mechanism of the homogeneous shear flow.
If $\tau_C$ is almost independent or, at least,  weakly dependent on $L_z$, the dynamical 
behavior of the system may be considered as substantially unaffected by the finite size of 
the box. 
A rigorous analysis of the correlation time dependence on the spanwise size $L_z$ would 
require a large number of highly resolved numerical simulations far beyond the present 
computer capabilities. In figure 13, we show a preliminary evidence that 
the behavior of the kinetic energy is not too sensitive to $L_z$, by comparing the 
simulations performed with $L_z= 2 \pi$ and $L_z= 4 \pi$.
The characteristic time of the generation mechanism of large scale vorticity seems not to 
depend on $L_z$. These results suggest that the finite size of the box does not affect 
in a significant way the dynamics and the statistical properties of the homogeneous 
shear flow, despite the importance of the basic mode, whose length scale coincides with 
that of the box, in the dynamics of the flow.
\section{Intermittency and scaling}
In the previous sections we have analyzed the influence of the shear on the dynamics of 
the vortical structures. We are now ready to discuss its effect on statistical features of 
turbulent fluctuations, such as scaling laws and intermittency.

Concerning the velocity field, a quantitative description can be given in terms of the
possible scaling behavior of the structure functions, i.e.  the moments of longitudinal 
velocity increments
\be
\label{strut_fun}
< \dV^p> = < \{ [ \vec u(\vec x + \vec r,t)- \vec u(\vec x,t)] \cdot 
\frac{\vec r}{r} \}^p >
\ee
where angular brackets denote ensemble averaging. 
For homogeneous and isotropic conditions, the dimensional prediction of Kolmogorov 
theory (K41) provides a scaling law in terms of separation \cite{kolm_41}
\begin{eqnarray}
\label{K41}
< \dV^p > & \propto & \bar{\epsilon}^{p/3} \, r^{p/3} \,
\end{eqnarray}
where $\bar \epsilon$ denotes the mean rate of energy dissipation per unit mass.
For $p=3$, 
equation (\ref{K41}) is consistent with the exact result usually referred to as 
the {\em{four fifth law}},
\begin{eqnarray}
\label{four_fifth}
< \dV^3 > & = & -\frac{4}{5} \bar{\epsilon} \, r \,.
\end{eqnarray}
This  equation has been widely confirmed by a number of experimental investigations 
\cite{benzi_1} which, at the same time, showed that dimensional scaling does not hold for 
moments different from the third. In fact, a more complex dependence of the scaling exponents 
is found,
\begin{eqnarray}
\label{EXP}
< \dV^p > & \propto &  r^{\zeta_p},
\end{eqnarray}
with $\zeta_p$ a nonlinear convex function of $p$. This anomalous scaling related to 
{\em{intermittency}}, has been actively investigated over the last twenty years. 

\subsection{Dimensionless parameters}
To understand how intermittent fluctuations are affected by the shear, it is worthwhile 
to consider the typical length scales involved in the process and the relevance
of the shear production term over the nonlinear inertial interactions. 
In presence of shear, velocity fluctuations arise for two main reasons, advection of
streamwise momentum across the mean gradient and non-linear mixing. Let us denote by 
$\delta U = S r$ the velocity difference at scale $r$ due to the mean flow and by 
$\delta u$ the fluctuation at the same scale. A rough estimate for the latter is 
$\delta u \propto \bar{\epsilon}^{1/3} r^{1/3}$.
The length scale for which turbulent fluctuations are of the same order of magnitude as
the velocity increments induced by the shear, uniquely identifies the {\sl shear scale} 
$L_s$ 
\be
\label{shear_scale}
L_s=\sqrt{\frac{\bar{\epsilon}}{S^3}} \, ,
\ee
that separates two different sub-ranges within the inertial range.
For $ L_0 \gg r \gg L_s$ (with $L_0$ the integral scale),  $\delta U \gg \delta u$ and
the dynamics and statistics of turbulence is expected to be dominated by the momentum flux 
due to the Reynolds stresses, i.e. by the production of kinetic energy.
For $\eta \ll r \ll L_s$, $\delta u \gg \delta U$ and the fluctuations induced by the mean 
shear are negligible with respect to those typically produced by the nonlinear term. 
In this case the dynamics of the flow is characterized by the energy transfer due to the same 
inertial interactions which are typical of isotropic turbulence.

In order to quantify the predominance of either mechanism over the other, we consider the
 dimensionless parameter, defined as the ratio of an inertial and a shear time scale,
\be
\label{shear_par1}
S^*=\frac{S q^2}{\bar{\epsilon}}
\ee
with $q^2=<u_i u_i>$. $S^*$ can be interpreted as the ratio of two length scales  
\be
\label{shear_par2}
S^*=\left(\frac{l_d}{L_s}\right)^{2/3} \,
\ee
where $l_d=q^3 / \bar{\epsilon}$. It follows that for large values of $S^*$ the dynamics 
of most of the inertial scales is dominated by the mean shear. 

The other dimensionless parameter needed for a complete description of a homogeneous
turbulent shear flow can be given as the ratio of the dissipative and the shear time scale,
\be
\label{shear_par3}
S^*_c=S (\nu / \bar{\epsilon})^{1/2} \, ,
\ee
or, equivalently, as the ratio of the Kolmogorov length and the shear scale
\be
\label{shear_par4}
S^*_c=\left(\frac{\eta}{L_s}\right)^{2/3}.
\ee
Since $S^*_c$ measures the separation between the shear scale and the dissipative range, 
a sub-range of isotropic behavior may be recovered within the inertial range when it is small.

Typical values of $S^*$ and $S^*_c$ are shown in figure 14, which is a
compilation of several independent numerical and experimental simulations.
The variety of behaviors observed in these simulations can easily be interpreted in terms of 
the location of the shear scale $L_s$, quite different from one case to the other. 

\subsection{Intermittency}
Let us now turn to the issue of intermittency. A quantitative measure of intermittency 
is provided by the flatness factor of the random variable $\dV(r)$. 
Figures 15 and 16 show the probability distribution function of 
$\dV(r)$ for 
two different separations both in the shear flow and  in homogeneous isotropic 
turbulence. 
For small separations in both cases the pdf is clearly non Gaussian and the flatness factor, 
shown in figure 17, grows as separation decreases confirming that 
the dimensional scaling of the structure functions is clearly violated. 
However, important differences are apparent.
The tails for the shear flow are quite higher and the pdf is more skewed towards negative 
values. 
For the larger separation, instead, the two pdfs are much more similar, though still not 
Gaussian. 
The Gaussian behavior is recovered only for $r$ comparable with the integral scale. 
Globally, we find a correspondence with what was found in the turbulent channel flow 
\cite{PF}, where at small separation the pdfs of $\dV(r)$ in the buffer and in the log-region 
differs substantially, with the negative tail much higher than the positive and an increased
flatness factor in the buffer.

\subsection{Similarity laws}
The issue of intermittency can be quantified by looking directly at the scaling properties of 
the structure functions. As well known, starting from Landau objection \cite{frisch} 
a revised form of similarity law has been proposed by Kolmogorov and Obhukov 
\cite{kolm_62} (K62): assuming that the dissipation field is a random variable, 
equation (\ref{K41}) is corrected to
\begin{eqnarray}
\label{K62}
< \dV^p > & \propto & <\epsilon_r^{p/3}> \, r^{p/3} \,,
\end{eqnarray}
where $<\epsilon_r^q>$ denotes the q-th moment of the local energy dissipation field 
$\epsilon_{loc}$ averaged over a volume of characteristic dimension $r$. 
Given for the dissipation field a scaling law in terms of separation,
\begin{eqnarray}
\label{eps_scaling}
< \epsilon_r^q > & \propto & r^{\tau(q)} \,,
\end{eqnarray}
from equation (\ref{K62}) the scaling exponents of the structure 
functions are expressed as
\be
\label{zeta_p}
\zeta(p) = \tau(p/3) + p/3
\ee
with $\tau(p/3)$ a nonlinear function of $p$ and $\tau(1)=0$.
Equation (\ref{zeta_p}) is consistent with the intermittent nature of the random variable 
$\dV(r)$, i.e. its increasing flatness with decreasing separation,
since the flatness factor, within the inertial range, turns out to be an unbounded function, 
diverging for small separations as $r^{\zeta_4-2\zeta_2}$.  
The intermittency is related to bursty signals characterized by localized
and intense fluctuations and to the presence of coherent structures and associated
regions of intense gradients which give rise to a highly non uniform dissipation field.  

Scaling laws in terms of separation  are found only for sufficiently large values 
of the Reynolds number i.e., so far, only in experimental facilities. 

An extension of the scaling range can be achieved by using the Extended Self Similarity (ESS) 
as proposed by Benzi et. al. \cite{benzi_1}, who introduced 
a generalized form of the scaling law (\ref{EXP}),
\begin{eqnarray}
\label{ESS}
< \dV^p > & \propto &   < \dV^3 >^{\zeta_p/\zeta_3} \,.
\end{eqnarray}
Using experimental data, the scaling law (\ref{ESS}) has been shown to hold even at low and 
moderate Reynolds number, i.e. in cases, such as those amenable to DNS,
where the scaling (\ref{EXP}) is not directly detectable. 
Most interestingly, the ratios $\zeta_p/\zeta_3$ have systematically been found independent
of the Reynolds number and consistent with the measurements in the high Reynolds number regime. 
The ESS allows a generalization of the refined Kolmogorov similarity hypothesis 
\cite{benzi_2}, eq. (\ref{K62}), namely
\begin{eqnarray}
\label{GESS}
< \dV^p > & \propto & \frac{< \epsilon_r^{p/3} >} {< \epsilon >^{p/3}}
\, < \dV^3 >^{p/3} \,.
\end{eqnarray}
This equation, which we will hereafter refer to as RKSH, has been carefully 
checked against experimental data for homogeneous and isotropic turbulence and confirmed also 
in those cases where neither $< \epsilon_r^{p/3}>$ nor the structure functions showed a 
scaling behavior in $r$. 
The two terms in eq.~(\ref{GESS}), namely $< \epsilon_r^{p/3} >$ and $< \dV^3 >$, take into 
account the fluctuations in the energy dissipation rate and the energy transfer through the 
inertial scales, respectively.

Scaling laws as discussed so far heavily rely on the assumption of isotropy.
In many turbulent flows, however, isotropy is broken by the mean shear and by the boundary 
conditions and a large production of turbulent kinetic energy occurs due to the interaction 
of the mean flow with the turbulent fluctuations.
We may wonder if scaling laws emerge also in such conditions, and, in case, under which
respects they differ from the classical ones.

Recent numerical investigations of wall turbulence have shown how intermittency increases
approaching the wall and how the scaling exponents $\zeta_p$ are not consistent with
equation (\ref{GESS}). Actually, in the wall region, the main dynamical process is 
represented by the momentum transfer occurring in the wall normal direction associated 
with a large production of turbulent kinetic energy via the Reynolds stresses
$d U /dy <u v>$. In this case a term of the form $d U /dy <\dV^2>$ is expected to
account for the new balance between production and dissipation in the Karman-Howarth 
equation. 
Based on this analysis a new form of similarity law has been proposed for the wall region 
in terms of the structure function of order two \cite{PF}
\begin{eqnarray}
\label{new}
< \dV^p > & \propto & \frac{< \epsilon_r^{p/2} >} {< \epsilon >^{p/2}}
\, < \dV^2 >^{p/2} \,.
\end{eqnarray}

As discussed in \cite{PF} for a turbulent channel flow, the classical RKSH, eq. (\ref{GESS}),
holds at the center of the channel in the bulk region where the momentum transfer 
due to the shear term is negligible with respect to inertial transfer. The new form of 
similarity law holds in the near wall region where high turbulent kinetic 
energy production occurs. 

We are presently going to check the consistency of equation (\ref{new}) in the homogeneous 
shear flow where, in the absence of solid walls, the mean shear $S$ is constant. 
In this flow the effect of the shear is isolated from other concurrent effects such as the 
suppression of wall normal velocity fluctuations and the non uniform momentum transfer across 
adjacent layers in the normal direction. Moreover the statistical analysis is simplified 
since homogeneity is exploited in all spatial directions.

In figure 18 the similarity law (\ref{GESS}), for $p=6$, is applied to DNS data 
for both isotropic and homogeneous shear turbulence. It holds for all the resolved 
scales in the isotropic case. For the shear flow, a clear failure is observed instead. 
In this case two different slopes may be extracted from the data, one in the 
dissipative region with the trivial value of $0.99$, the other, in the inertial region, 
with value of $0.92$.  
Analogous results are found for $p=9$, figure 19, with a single slope for 
isotropic turbulence.
For the shear flow, the dissipative slope is $0.98$ confirming the quality of our DNS
while the inertial fit gives a slope of $0.88$.

A similar analysis is presented in figures 20 and 21 
for the new similarity law,
eq. (\ref{new}), in the case of the homogeneous shear flow,  for $p=4$ and $p=6$. 
A unique slope can be effectively fitted all over the range of resolved scales with the same 
value of $1$ independently of the order of the moment. 
The failure of the classical RKSH and the validity of the new form for the shear flow
are still better appreciated in the compensated plot of equations (\ref{GESS}) and (\ref{new}) 
versus the separation $r$, figure 22 and 23.

As a technical detail concerning the evaluation of the RKSH, we remark
that $L_s=0.8$ and that the dissipation fit is extracted from the range of separations 
$[0, \, l_{fit}]$, with $l_{fit}=0.2$. 
According to our theoretical considerations, for $r \ll L_s$ we should recover the classical 
RKSH. 
However, the present value of $0.15$ for $S_c^*$ implies that when $r \ll L_s$ we also 
have $r \sim \eta$, hence no room is left for the establishment of the classical form 
of scaling below $L_s$.

\subsection{Extended self similarity}
Before closing the section, it is  worth commenting on the issue of ESS scaling in 
shear flows.
Actually, the use of equation (\ref{ESS}) requires some caution in the present conditions, 
where a superposition of different laws may be present.
Clearly, a blind fit through the entire available range can extract an effective slope. 
It is much safer, instead, to consider the local slopes in detail, to understand to what 
extent a power law is really observed, and, more interestingly, the range of scales over which 
a certain value for the exponent is found.
As an example, in figures 24 and 25,
the local slopes,  $p=6$ and $p=9$, for the shear flow are contrasted with those of  
homogeneous isotropic turbulence.
In both cases at small separation the dissipative scaling $\dV^p \propto r^p$ is 
achieved confirming the good quality of the data. As we move towards inertial separations
the local slopes of isotropic turbulence remain quite constant over almost one decade.
For the shear flow instead a definite deviation from constant is found at approximately 
$r / \eta =15$, with the shear scale falling near by ($L_s/\eta=16$). 
Similarly to what found in the lower part of the logarithmic region of the turbulent 
channel \cite{PRL}, the local slopes do not display clear ESS scaling regions, presumably 
due to the competition of two different statistical trends and to the lack of sufficient 
scale separation to distinguish between the two.

\subsection{An estimate of the flatness}
A different estimate of the flatness is here proposed taking into account the similarity
laws given by eq. (\ref{GESS}) and (\ref{new}). Coming back to the issue of 
intermittency, let us consider again the flatness factor of the random variable 
$\dV(r)$, 
\begin{eqnarray}
\label{F_dv}
F(r) & = &  \frac{<\dV^4(r)>}{<\dV^2(r)>^2}  \, .
\end{eqnarray}
According to the two forms of RKSH, eqs.~(\ref{GESS}) and (\ref{new}), the same flatness 
factor can be expressed in terms of moments of the dissipation field as
\begin{eqnarray}
\label{F_w}
F_b  \, = \, \frac{<\epsilon_r^{4/3}>}{<\epsilon_r^{2/3}>^2} & \qquad \qquad &
F_w  \, = \, \frac{<\epsilon_r^{2}>}{<\epsilon_r>^2}  \,,
\end{eqnarray}
respectively.
The plots in figure 26 show the curves corresponding to the 
three equations (\ref{F_dv}) and (\ref{F_w}). 
The values computed using the new RKSH, $F_w$ in eq.~(\ref{F_w}), are in excellent
agreement with those evaluated from the definition (\ref{F_dv}). 
It should be noted that the use of the wrong form of RKSH, corresponding to expression
$F_b$ in eq.~(\ref{F_w}), substantially underestimates the intermittency.

The same analysis can be performed for homogeneous and isotropic turbulence. In the plots
of figure 27, we show $F$, $F_b$ and $F_w$ for turbulence 
measured at the center of a jet far downstream ($R_\lambda \sim 800$)\cite{benzi_1}. 
As one can easily observe, the new form of refined similarity hypothesis fails in homogeneous 
isotropic conditions, as we should expect.

The ability of the new RKSH to predict the amount of intermittency
in the velocity increments from the pdf of the dissipation field is, in our opinion, 
a strong check on the validity of the proposed form of scaling.
\section{Statistical properties of the dissipation field}

Given its role in both the new and the classical form of RKSH, the study of the dissipation 
field may give further insight in the behavior of shear turbulence. To this purpose 
figure 28 reports the self-scaling of the dissipation field
in logarithmic scale. The solid line with slope $\alpha =3.1$ shows clearly that for 
sufficiently large separations $r$, which  correspond to small values of the two moments, 
the power law $<\epsilon_r^3> \; \propto \; <\epsilon_r^2>^\alpha$ is able to fit the data.
The scaling is strikingly similar to that found in homogenous and isotropic turbulence, 
where the estimated exponent is $3$.
The self-scaling seems to hold for all the moments we have been able to check, and
the values found for the shear flow are always in agreement with those of homogeneous 
and isotropic turbulence, see table \ref{tabella}.
This analysis suggests that relations of the form 
\begin{eqnarray}
\label{diss_scal}
<\epsilon_r^q> & \propto & <\epsilon_r^2>^{\alpha_q}
\end{eqnarray}
\noindent with, more or less, fixed exponents are able to describe the statistics
of the fluctuations of the dissipation field either in shear or isotropic turbulence.
Hence, basic statistical properties of the dissipation field remain substantially unchanged
in the two cases.  This issue becomes even more evident when considering the second moment 
of the dissipation as a function of $r$. This quantity  is reported in figure 
29, where the triangles and the circles, corresponding to isotropic 
and shear turbulence respectively, indicate a substantially identical behavior.
All together, these findings suggest a universal form for the pdf of the dissipation field, 
$p(\epsilon_r / \bar{\epsilon})$ normalized by its mean value $\bar{\epsilon}$ which is
obviously dependent on the flow field.
\section{Concluding remarks \& comments}

The direct numerical simulation of a homogeneous shear flow in a confined box has been 
analyzed to address the issue of scaling laws and intermittency in the simplest conceivable 
environment where anisotropy is prevailing. The dynamics of the vortical structures observed 
in the DNS is characterized by a regeneration cycle associated to a substantial fluctuation 
in the level of turbulent kinetic energy. The regeneration mechanism has been discussed in 
detail, and our analysis confirms substantially to previous descriptions \cite{kida} which, 
however, where restricted to the initial phase of flow development.
In fact, in complete agreement with \cite{pumir_2} and \cite{pumir_3}, we confirm that 
the confinement of the
flow to a periodic box generates a dynamics, which, although strongly characterized by the
regeneration cycle of the vortical structures, reaches statically stationary conditions. 
The characteristic time of the system seems to be quite insensitive to changes in the aspect 
ratio of the computational box,  hence we feel that the present flow may be considered as a 
valid prototype for the study of shear flows. 

We find an increased intermittency, as evaluated by the flatness of the velocity increments,
with respect to that observed in isotropic turbulence. In isotropic turbulence, the 
intermittency of the velocity increments is related to the intermittent behavior of the 
dissipation field via the Refined Kolmogorov Similarity Hypothesis.
For the shear flow, instead, we observe a clear failure of the RKSH which confirms the 
conclusions previously reached from data in the buffer layer of a channel flow (DNS) and 
successively verified with hot-wire measurements in a flat plate boundary layer.

To attempt a clarification of this point, we have considered the possible existence of a 
self-scaling behavior of the structure functions.
Under this respect, the extended self-similarity expressing the $p^{th}$-order 
structure function as a power law with exponent $\zeta(p)$ in terms of, say, the third 
order moment, is a well established feature of isotropic turbulence which is able
to characterize the amount of intermittency in terms of the scaling exponents.
For the homogeneous shear flow, as already observed in the lower part of the logarithmic region of 
turbulent channel flow, scaling exponents can be defined only for a limited region 
of scales ($r < L_s$) above which a competition between two different statistical behaviors 
makes it difficult to define any scaling at all. However, recent experimental 
results concerning the log-layer of a turbulent boundary layer at a higher Reynolds
number, seem to indicate the existence of a double scaling regime \cite{ruiz}.

Nonetheless, we are presently able to grasp the nature of the intermittency in the shear flow,
since  the failure of the RKSH is accompanied by the establishment of a new form of 
scaling, as proposed in \cite{PF}.
Originally the idea behind the new scaling law was to account for the crucial role of the 
shear in originating a substantial production of turbulent kinetic energy via the Reynolds 
stress.  
In principle, this mechanism, directly associated to the presence of a mean shear,
is independent of other details of the flow, such as the presence of nearby
solid boundary or the specific dynamics of the vortical structures which sustain the turbulent 
fluctuations. This view is confirmed by the present results which show how the new form
of RKSH holds not only for the buffer layer of wall bounded flows, as originally verified in 
\cite{PF}, but also in the present homogeneous shear flow.
Here the dynamics of the vortical structures, a part from a general resemblance, is in fact 
much different from that observed in the wall region and no wall is present to inhibit
wall normal motions. Nonetheless the new scaling law is well established, confirming that its
presence is the signature of a predominant effect of the shear in the proper range of scales.

Beyond the direct observation of the new scaling law, the detailed analysis presented in the
paper shows how the increased intermittency of the longitudinal velocity increments in shear 
dominated flow is, in fact, to be ascribed to the new scaling law.
Actually, the statistical properties of the energy dissipation field are substantially
similar to those we find in isotropic conditions.
From our data, the scaling behavior of the energy dissipation field can not be consistent 
both with the observed level of intermittency of the velocity structure functions and with
the existence of the classical RKSH.
On the contrary, the new RKSH is able to  consistently predict the level of intermittency in 
the velocity structure function in terms of the sole dissipation field.
This confirms that the intermittency of velocity structure functions at shear dominated scales
is largely affected by the instantaneous process of turbulent kinetic energy production. 
In other words, since the dissipation field is approximately unchanged, to compute the
increased intermittency of the velocity in presence of shear we have to include 
explicitly the production mechanism in the scaling law.

These comments suggest that the source of intermittency in the shear flow is in principle 
substantially different from that operating in isotropic turbulence.
To speculate a little further on this point, let us recall that, a part from the aspect ratio 
of the box, the confined homogeneous shear flow is defined by two dimensionless parameters, 
namely $S^*$ and $S_c^*$. 
These two parameters control the position of the shear scale $L_s$ relative to the length scale
of the large eddies, $l_d$, and to the Kolmogorov scale, $\eta$, respectively.
A large value of $S^*$ implies a substantial range of scales where the shear is crucial
to the dynamics. 
On the other hand, a small value of $S_c^*$  corresponds to an extended range of scales 
where the dynamics is purely inertial and where the effect of shear is largely irrelevant.
One should expect the classical scaling laws of homogeneous and isotropic turbulence 
to emerge clearly in the limit of $S_c^*$ small within the range $ L_s >> r >> \eta$.
A major contribution of the present paper is to confirm the existence of a new form of scaling
law which emerges in the limit of $S^*$ large, in the range $l_d >> r >> L_s$.
Ideally, when both $S^*$ and $1/S_c^*$ are large, one should expect a coexistence of the two
laws in the two different scaling ranges.
In this ideal conditions, the intermittent behavior of velocity increments is induced
by the energy injection associated to the turbulent kinetic energy production mechanism
and is originated in the shear dominated range, where the scaling is totally different from
the classical RKSH.
However, once the energy is transferred to the smaller scales where direct injection
due to shear is irrelevant, the classical energy cascade mechanism takes over.
The intermittency of the velocity increments at the shear dominated scales appears to smaller 
scales which belongs to the classical inertial range as large scale modulations. 
Along the classical line of reasoning, the large scale modulations, through the cascade, are 
reduced to the intermittent behavior  of the dissipation field.
This is tantamount to postulating a certain universality of the probability distribution of the 
dissipation field, once, in the spirit of K62 theory, universal values for the anomalous 
corrections to the exponents of the structure functions are assumed in the classical part of
the inertial range. 
These concepts are consistent with our numerical experiments with the homogeneous shear flow, 
even though the scale separation is not sufficient to observe both scalings simultaneously.

To conclude we like to comment about the different sets of data available to
us in the present analysis.  As figure \ref{mappa_any} shows, a large shear dominated 
range, i.e. $S^*$ large, is not frequently met in practice.
An exception is the buffer region of wall bounded flows. There, almost the entire range of
available scales falls above $L_s$, which for this case is also very close to the Kolmogorov 
length, $\eta$, since $S_c^* \simeq 1$. 
These are optimal conditions to observe the new RKSH as a pure law, without interference with 
its classical counterpart since the contiguous purely inertial range below $L_s$ does not exist
at all. The logarithmic region, both from DNS and experiments,  present  
$L_s \simeq l_d$, i.e. $S^* \simeq 1$, and $L_s >> \eta$. Here only the classical scaling 
can be expected. The conditions of the present confined homogeneous shear flows are somehow 
intermediate between the ones of the buffer and the log region respectively.
Here $L_s$ falls almost exactly in the middle of the available range of scales, a more complex
situation and by far the more interesting to discuss. If the shear dominated and the purely 
inertial range were large enough we might have observed the simultaneous presence of the two 
forms of scaling law. However this cannot be  the case with the Reynolds number we can reach 
by DNS. In practice, what we observe in the present case  can be described as a sort of 
superposition of the two regimes, with a dominating contribution from the new RKSH. 
\newpage
\begin{table}
\caption{Scaling exponents, $\alpha_q$, of the moments $<\epsilon_r^q>$ with respect to 
         $<\epsilon_r^2>$, for different, non-integer, values of $q$, 
         see eq. (\ref{diss_scal}).
         \label{tabella} }
\begin{tabular}{ccc}
 $\alpha_{q}$  &   Hom. Iso.  &      Hom. Shear      \\
  \hline
$\alpha_{2/3}$ &   -0.11      &      -0.11           \\
$\alpha_{4/3}$ &    0.22      &       0.22           \\
$\alpha_{5/3}$ &    0.55      &       0.55           \\
$\alpha_{6/3}$ &    1.00      &       1.00           \\
$\alpha_{7/3}$ &    1.56      &       1.57           \\
$\alpha_{8/3}$ &    2.10      &       2.25           \\
$\alpha_{9/3}$ &    3.00      &       3.10           \\
\end{tabular}
\end{table}

\end{document}